\documentclass[
    ,final            
  ]
  {aipproc}

\layoutstyle{6x9}

\usepackage{graphicx}
\usepackage{amsmath}
\usepackage{color}

\definecolor{darkgreen}{rgb}{0,0.5,0}
\definecolor{purple}{rgb}{0.5,0,0.5}
\definecolor{nblue}{rgb}{0.0,0.0,0.50}
\definecolor{scarlet}{rgb}{1.0,0.2,0}

\newcommand{\lsim}{\mathrel{\rlap{\lower4pt\hbox{\hskip0pt$\sim$}}
\raise1pt\hbox{$<$}}}           
\newcommand{\gsim}{\mathrel{\rlap{\lower4pt\hbox{\hskip0pt$\sim$}}
\raise1pt\hbox{$>$}}}           

\begin{document}

\title{T(r)opical Dyson-Schwinger Equations}


\author{L.~Chang$^{1,2}$, I.\,C.~Clo\"et$^{3,4}$,
C.\,D.~Roberts$^{1,2,4,5}$
and H.\,L.\,L.~Roberts$^{2,5,6}$}{
  address={
  $^1$Department of Physics, Center for High Energy Physics and the State Key Laboratory of Nuclear Physics and Technology, Peking University, Beijing 100871, China\\
  $^2$Physics Division, Argonne National Laboratory, Argonne, Illinois 60439, USA (http://www.phy.anl.gov/theory/index.html)\\
  $^3$Department of Physics, University of Washington, Seattle WA 98195, USA\\
  $^4$Kavli Institute for Theoretical Physics China, CAS, Beijing 100190, China\\
  $^5$Institut f\"ur Kernphysik, Forschungszentrum J\"ulich, D-52425 J\"ulich, Germany\\
  $^6$Physics Department, University of California, Berkeley, California 94720, USA
  }}


    %
    %

  %
  %
  %

\begin{abstract}
We provide a glimpse of recent progress in hadron physics made using QCD's Dyson-Schwinger equations, reviewing: the notion of in-hadron condensates and a putative solution of a gross problem with the cosmological constant; the dynamical generation of quark anomalous chromo- and electro-magnetic moments, and their material impact upon the proton's electric/magnetic form factor ratio; a computation that simultaneously correlates the masses of meson and baryon ground- and excited-states; and a prediction for the $x\to 1$ value of the ratio of neutron/proton distribution functions.
\end{abstract}

\maketitle

With QCD, Nature has given us the sole known example of a strongly-interacting quantum field theory that is defined through degrees-of-freedom which cannot directly be detected.  This empirical fact of \emph{confinement} ensures that QCD is the most interesting and challenging piece of the Standard Model.  It means that building a bridge between QCD and the observed properties of hadrons is one of the key problems for modern science.  In confronting this challenge, steps are being taken with approaches that can rigorously be connected with QCD.  For example, in the continuation of more-than thirty-years of effort, extensive resources are being spent on the application of numerical simulations of lattice-regularised QCD \cite{LAT2009}.  Herein, on the other hand, we recapitulate on efforts within the Dyson-Schwinger equation (DSE) framework \cite{Roberts:1994dr,Roberts:2000aa,Maris:2003vk,Chang:2010jq}, which provides a different, continuum perspective to computing hadron properties from QCD.

No solution to QCD will be complete if it does not explain confinement.  This means confinement in the real world, which contains quarks with light current-masses.  That is distinct from the artificial universe of pure-gauge QCD without dynamical quarks, studies of which tend merely to focus on achieving an area law for a Wilson loop and hence are irrelevant to the question of light-quark confinement.  Confinement can be related to the analytic properties of QCD's Schwinger functions \cite{Krein:1990sf,Roberts:2007ji} and may therefore be translated into the challenge of charting the infrared behavior of QCD's \emph{universal} $\beta$-function.  This is a well-posed problem whose solution can be addressed in any framework enabling the nonperturbative evaluation of renormalisation constants.

Dynamical chiral symmetry breaking (DCSB); namely, the generation of mass \emph{from nothing}, is a fact in QCD.  This is best seen by solving the DSE for the dressed-quark propagator \cite{Bhagwat:2003vw,Bhagwat:2006tu,Bhagwat:2007vx}; i.e., the gap equation.  However, the origin of the interaction strength at infrared momenta, which guarantees DCSB through the gap equation, is unknown.  This relationship ties confinement to DCSB.  The reality of DCSB means that the Higgs mechanism is largely irrelevant to the bulk of normal matter in the universe.  Instead the single most important mass generating mechanism for light-quark hadrons is the strong interaction effect of DCSB; e.g., one can identify it as being responsible for 98\% of a proton's mass.

We note that the hadron spectrum \cite{Holl:2005vu}, and hadron elastic and transition form factors \cite{Cloet:2008re,Aznauryan:2009da} provide unique information about the long-range interaction between light-quarks and, in addition, the distribution of a hadron's characterising properties amongst its QCD constituents.  However, to make full use of extant and forthcoming data, it will be necessary to have Poincar\'e covariant theoretical tools that enable the reliable study of hadrons in the mass range $1$-$2\,$GeV.  Crucially, on this domain both confinement and DCSB are germane; and the DSEs provide such a tool.

For the last thirty years, \emph{condensates}; i.e., nonzero vacuum expectation values of local operators, have been used as parameters in order to correlate and estimate essentially nonperturbative strong-interaction matrix elements.  They are also basic to
current algebra analyses.  It is conventionally held that such quark and gluon condensates have a physical existence, which is independent of the hadrons that express QCD's asymptotically realisable degrees-of-freedom; namely, that these condensates
are not merely mass-dimensioned parameters in a theoretical truncation scheme, but in fact describe measurable spacetime-independent configurations of QCD's elementary degrees-of-freedom in a hadronless ground state.

However, it has been argued that this view is erroneous owing to confinement  \cite{Brodsky:2010xf}.  Indeed, it was proven \cite{Maris:1997hd,Maris:1997tm} that the chiral-limit vacuum quark condensate is qualitatively equivalent to the pseudoscalar-meson leptonic decay constant in the sense that both are obtained as the chiral-limit value of well-defined gauge-invariant hadron-to-vacuum transition amplitudes that possess a spectral representation in terms of the current-quark-mass.  Thus, whereas it might sometimes be convenient to imagine otherwise, neither is essentially a constant mass-scale that fills all spacetime.  Hence, in particular, the quark condensate can be understood as a property of hadrons themselves, which is expressed, for example, in their Bethe-Salpeter or light-front wave functions.  In the latter instance, the light-front-instantaneous quark propagator appears to play a crucial role \cite{Brodsky:2010xf,Burkardt:1998dd}

This has enormous implications for the cosmological constant.  The universe is expanding at an ever-increasing rate and theoretical physics has tried to explain this in terms of the energy of quantum processes in vacuum; viz., condensates carry energy and so, if they exist, they must contribute to the universe's dark energy, which is expressed in the computed value of the cosmological constant.  The problem is that hitherto all potential sources of this so-called vacuum energy give magnitudes that far exceed the value of the cosmological constant that is empirically determined.  This has been described as ``the biggest embarrassment in theoretical physics'' \cite{Turner:2001yu}.  However, given that in the presence of confinement condensates do not leak from within hadrons, then there are no space-time-independent condensates permeating the universe \cite{Brodsky:2010xf}.  This nullifies completely their contribution to the cosmological constant and reduces the mismatch between theory and observation by a factor of $10^{46}$ \cite{Brodsky:2009zd}, and possibly by far more, if technicolour-like theories are the correct paradigm for extending the Standard Model.

In QCD, DCSB is most basically expressed through a strongly momentum-dependent dressed-quark mass; viz., $M(p^2)$ in the quark propagator:
\begin{equation}
S(p) = \frac{1}{i \gamma\cdot p A(p^2) + B(p^2)} = \frac{Z(p^2)}{i \gamma\cdot p + M(p^2)}\,.
\end{equation}
The appearance and behaviour of $M(p^2)$ are essentially quantum field theoretic effects, unrealisable in quantum mechanics.  The running mass connects the infrared and ultraviolet regimes of the theory, and establishes that the constituent-quark and current-quark masses are simply two connected points on a single curve separated by a large momentum interval.  QCD's dressed-quark behaves as a constituent-quark, a current-quark, or something in between, depending on the momentum of the probe which explores the bound-state containing the dressed-quark.  These remarks should make clear that QCD's dressed-quarks are not simply Dirac particles, a fact which can be elucidated further using a novel formulation of the Bethe-Salpeter equation \cite{Chang:2009zb,Chang:2010hb}.

In Dirac's relativistic quantum mechanics, a fermion with charge $q$ and mass $m$, interacting with an electromagnetic
field, has a magnetic moment $\mu = q/[2 m]$.  This prediction held true for the electron until improvements in experimental techniques enabled the discovery of a small deviation.  This correction was explained by the first systematic computation using renormalised quantum electrodynamics (QED) \cite{Schwinger:1948iu}:
\begin{equation}
\label{anommme}
\frac{q}{2m} \to \left(1 + \frac{\alpha}{2\pi}\right) \frac{q}{2m}\,,
\end{equation}
where $\alpha$ is QED's fine structure constant.  In general the current describing a photon-fermion interaction can be written
\begin{equation}
i q \, \bar u(p_f) \left[ \gamma_\mu F_1(k^2)+ \frac{1}{2 m} \sigma_{\mu\nu} k_\nu F_2(k^2)\right] u(p_i)\,,
\end{equation}
where: $F_1(k^2)$, $F_2(k^2)$ are form factors and $u(p)$, $\bar u(p)$ are spinors.  Dirac's result corresponds to $F_1\equiv 1$, $F_2\equiv 0$ whereas the anomalous magnetic moment in Eq.\,(\ref{anommme}) corresponds to $F_2(0) = \alpha/2\pi$.

It is a fascinating feature of quantum field theory that a massless fermion does not possess a measurable magnetic moment.  This fact is not in conflict with Eq.\,(\ref{anommme}) because the perturbative expression for $F_2(0)$ contains a multiplicative numerator factor of $m$ and the usual analysis of the denominator involves steps that are only valid for $m\neq 0$.  The analogue of Schwinger's one-loop calculation can be carried out in QCD to find an anomalous \emph{chromo}-magnetic moment for the quark.  There are two diagrams in this case: one similar in form to that in QED; and another owing to the gluon self-interaction.  One reads from Ref.\,\cite{Davydychev:2000rt} that the contribution from the three-gluon-vertex diagram vanishes identically and the remaining term is zero in the chiral limit.

The situation changes dramatically when chiral symmetry is dynamically broken.  This produces a dressed light-quark with a momentum-dependent anomalous chromomagnetic moment, which is large at infrared momenta and whose existence is likely to have numerous observable consequences \cite{Chang:2010jq,Kochelev:1996pv,Diakonov:2002fq,Ebert:2005es}.  Significant amongst them is the generation of an anomalous electromagnetic moment for the dressed light-quark with commensurate size but opposite sign \cite{Chang:2010hb}.  The infrared scale of both dynamically-generated moments is determined by the Euclidean constituent-quark mass \cite{Maris:1997tm}.  This is two orders-of-magnitude greater than the physical light-quark current-mass, which sets the scale of the perturbative result for both these quantities.  There are two additional notable and model-independent features; namely, the rainbow-ladder truncation, and low-order stepwise improvements thereof, underestimate these effects by an order of magnitude; and both the so-called $\tau_4$ and $\tau_5$ terms in the dressed-quark-gluon vertex are indispensable for a realistic description of hadron phenomena.

\begin{figure}[t]
\centerline{\includegraphics[clip,width=0.73\textwidth]{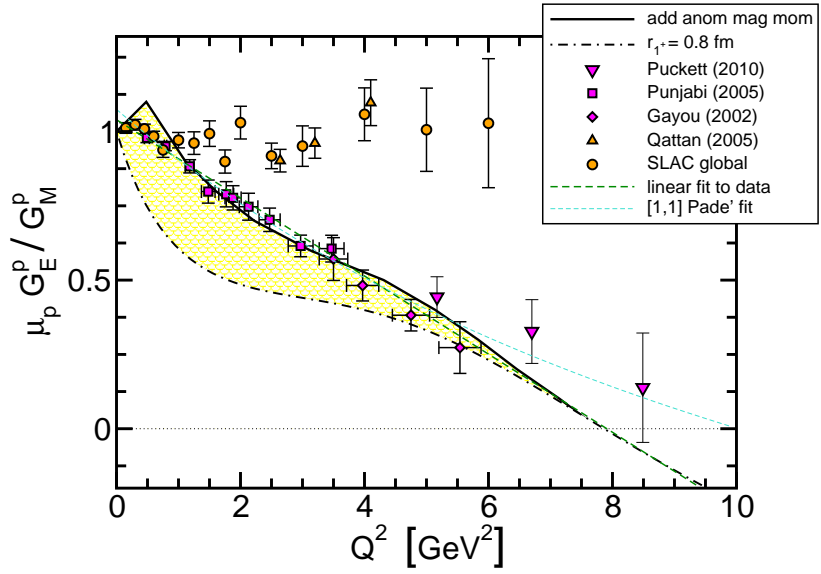}}
\caption{\label{ICCloetFig}
Proton electric-to-magnetic form factor ratio.  The \emph{dot-dashed curve} is the result in Ref.\,\protect\cite{Cloet:2008re}, whereas the \emph{solid curve} is obtained by repeating that calculation with inclusion of a momentum-dependent dressed-quark anomalous magnetic moment that is characterised by a $Q^2=0$ strength $\eta_{\rm em}=0.4$.  Data: diamonds -- \protect\cite{gayou}; squares -- \protect\cite{Punjabi:2005wq}; up-triangles -- \protect\cite{Qattan:2004ht}; circles \protect\cite{Walker:1993vj}; and down-triangles \protect\cite{Puckett:2010ac}.
\emph{Dashed curve}: $[1,1]$-Pad\'e fit to available JLab data; and \emph{dotted curve}, a linear fit.}
\end{figure}

The dynamical generation of dressed-quark anomalous magnetic moments is a fundamentally new phenomenon in quantum field theory.  Its discovery is stimulating a reanalysis of the hadron spectrum and hadron elastic and transition electromagnetic form factors.  Indeed computations are already underway, which elucidate the contribution of the dressed-quark anomalous magnetic moment to the nucleons' magnetic moments and electromagnetic form factors.   A preliminary result from this study is displayed in Fig.\,\ref{ICCloetFig}.  The effect on the ratio is large at infrared momenta.  Given years of community experience with constituent-quark-models \cite{Chung:1991st,Capstick:1994ne,Cardarelli:1995dc}, this had to be expected once a sound foundation was laid for nonzero dressed-quark anomalous magnetic moments.  On the other hand, the effect diminishes rapidly with increasing momentum transfer owing to the essential momentum-dependence of the dressed-quark anomalous magnetic moment.

Elucidation of the connection between DCSB and dressed-quark anomalous magnetic moments was made possible by the derivation of a novel form for the relativistic bound-state equation \cite{Chang:2009zb}, which is valid and tractable when the quark-gluon vertex is fully dressed.  This has also enabled an exposition of the impact of DCSB on the hadron spectrum.  For example, spin-orbit splitting between ground-state mesons is dramatically enhanced and this is the mechanism responsible for a magnified splitting between parity partners; namely, essentially-nonperturbative DCSB corrections to the rainbow-ladder truncation\footnote{This is the leading-order in a nonperturbative, systematic and symmetry-preserving DSE truncation scheme \protect\cite{Munczek:1994zz,Bender:1996bb}.} largely-cancel in the pseudoscalar and vector channels \cite{Bhagwat:2004hn} but add constructively in the scalar and axial-vector channels \cite{Chang:2010jq}.

\begin{figure}[t]
\includegraphics[clip,width=0.73\textwidth]{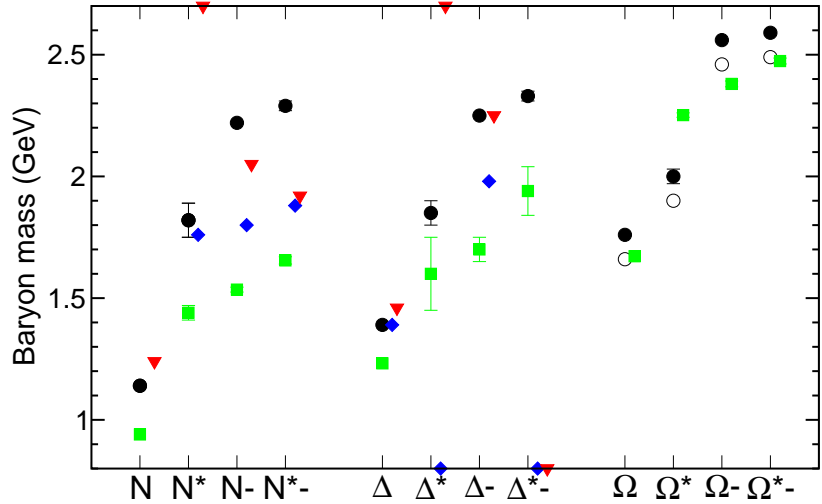}
\caption{\label{Fig2}
Comparison between DSE-computed baryon masses (\emph{filled circles}) and bare masses determined in Ref.\,\protect\cite{Suzuki:2009nj} (\emph{filled diamonds}) and Ref.\,\protect\cite{Gasparyan:2003fp} (\emph{filled triangles}).
For the coupled-channels models a symbol at the lower extremity indicates that no associated state is found in the analysis, whilst a symbol at the upper extremity indicates that the analysis reports a dynamically-generated resonance with no corresponding bare-baryon state.
In connection with $\Omega$-baryons the \emph{open-circles} represent a shift downward in the computed results by $100\,$MeV.  [This is an estimate of the effect produced by pseudoscalar-meson loop corrections in $\Delta$-like systems at a $s$-quark current-mass.]
The \emph{filled-squares} report masses in Ref.\,\protect\cite{Nakamura:2010zzi}.
}
\end{figure}

Building on such results within the context of a symmetry-preserving regularisation of the vector$\,\times\,$vector contact interaction \cite{Roberts:2010rn}, it is now possible to complete a Dyson-Schwinger equation calculation of the light hadron spectrum that simultaneously correlates the masses of meson and baryon ground- and excited-states within a single framework.  In comparison with relevant quantities the computation produces $\overline{\mbox{rms}}$=13\%, where $\overline{\mbox{rms}}$ is the root-mean-square-relative-error$/$degree-of freedom.  Notable amongst the results is agreement between the computed baryon masses and the bare masses employed in modern dynamical coupled-channels models of pion-nucleon reactions.  This is illustrated in Fig.\,\ref{Fig2}.  The analysis also provides insight into numerous aspects of baryon structure; e.g., relationships between the nucleon and $\Delta$ masses and those of the dressed-quark and diquark correlations they contain.

\begin{figure}[t]
\hspace*{6em}\includegraphics[clip,width=0.73\textwidth]{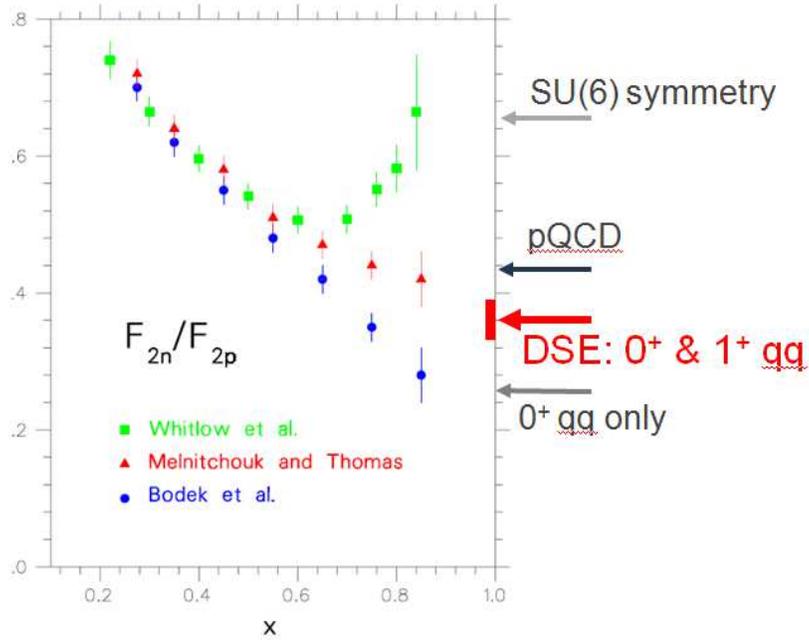}
\caption{\label{Fig3}
Prediction of the $x=1$ value for the ratio of neutron-to-proton structure functions based on the calculations of Ref.\,\protect\cite{Cloet:2008re} cf.\ $SU(6)$-symmetric constituent-quark-model, pQCD and the result obtained if the nucleon's Faddeev amplitude contains no axial-vector diquark correlation.
Extractions from data:
squares -- \protect\cite{Whitlow:1990gk};
triangles -- \protect\cite{Thomas:1997rp};
and circles \protect\cite{Whitlow:1991uw}.
(For a more complete explanation, see Ref.\,\protect\cite{Holt:2010vj}.)
}
\end{figure}
This relative strength of the axial-vector diquark cf.\ the scalar diquark has a big impact on nucleon properties.  For example, in the electromagnetic form factor calculations described in Ref.\,\cite{Cloet:2008re}, the photon-nucleon interaction involves an axial-vector diquark correlation with 40\% probability.  This value: is crucial in fixing the location of a zero in the ratio $F_1^{p,d}(Q^2)/F_1^{p,u}(Q^2)$ \cite{Roberts:2010hu}; determines the $x=1$ value for ratio of nucleon structure functions $F_2^n/F_2^p = 0.36\pm 0.03$ \cite{Holt:2010vj} see Fig.\,\ref{Fig3}; and entails that in the nucleon's rest frame just 37\% of the total spin of the nucleon is contained within components of the Faddeev amplitude which possess zero quark orbital angular momentum \cite{Cloet:2007pi}.

There are many reasons why this is an exciting time in hadron physics.  We have focused on one.  Namely, through the DSEs, we are positioned to unify phenomena as apparently diverse as the: hadron spectrum; hadron elastic and transition form factors, from small- to large-$Q^2$; and parton distribution functions.  The key is an understanding of both the fundamental origin of nuclear mass and the far-reaching consequences of the mechanism responsible; namely, DCSB.  These things might lead us to an explanation of confinement, the phenomenon that makes nonperturbative QCD the most interesting piece of the Standard Model.



\bigskip

\hspace*{-\parindent}\mbox{\textbf{Acknowledgments.}}~We acknowledge valuable discussions with M.~D\"oring, S.~Krewald, T.\,S-H.~Lee, C.~Hanhart and S.\,M.~Schmidt.
Work supported by:
Forschungszentrum J\"ulich GmbH;
the U.\,S.\ Department of Energy, Office of Nuclear Physics, contract nos.~DE-FG03-97ER4014 and DE-AC02-06CH11357;
and the Department of Energy's Science Undergraduate Laboratory Internship programme.

\vspace*{-2ex}



\bibliographystyle{aipproc}   


\IfFileExists{\jobname.bbl}{}
 {\typeout{}
  \typeout{******************************************}
  \typeout{** Please run "bibtex \jobname" to optain}
  \typeout{** the bibliography and then re-run LaTeX}
  \typeout{** twice to fix the references!}
  \typeout{******************************************}
  \typeout{}
 }


\end{document}